# Use of Blended Approach in the Learning of Electromagnetic Induction

Charles **CHEW**
*Principal Master Teacher*
*Academy of Singapore Teachers*
*Ministry of Education, Singapore*
Email : Charles_CHEW@moe.gov.sg

Loo Kang **WEE**
*Senior Specialist, Media Design & Technologies for Learning Branch*
*Education Technology Division*
*Ministry of Education, Singapore*
Email : WEE_Loo_Kang@moe.gov.sg

**ABSTRACT**

This paper traces the importance of pedagogical content knowledge in the digital age to prepare today's students for the 21$^{st}$ century. It highlights the need for ICT-based pedagogical models that are grounded in both the learning theories of constructivism and connectivism. One such suitable ICT-based pedagogical model is the TSOI Hybrid Learning Model. By means of a physics blended learning exemplar based on the TSOI Hybrid Learning Model, this paper argues for the use of blended learning approach as the way forward for 21$^{st}$ century teaching.

**INTRODUCTION**

In today's 21$^{st}$ century knowledge-based economy driven by the twin forces of globalization and the relentless advancements in Information Communication Technology (ICT), governments around the world are acutely aware of the vital role of schools in the success of both individuals and nations. This vital role of schools in equipping students with greater knowledge, higher-order thinking and performance skills to survive and succeed in the 21$^{st}$ century is highlighted by the chairman of the board of directors and chief executive officer of Dell Inc, Mr Michael Dell:

> *Reading, math and science are the foundations of student achievement. But to compete and win in the global economy, today's students and tomorrow's leaders need another set of knowledge and skills. These 21$^{st}$ century skills include the development of global awareness and the ability to collaborate and communicate and analyse and address problems. And they need to rely on critical thinking and problem solving to create innovative solutions to the issues facing our world. Every child should have the opportunity to acquire and master these skills and our schools play a vital role in making this happen.*
>
> *(Dell, 2009, p. 4)*

If we acknowledge that schools play a vital role in equipping students with the 21st century knowledge, skills and values, quality school leadership, systems and educational programs are necessary but not sufficient conditions. For any educational program to be successfully implemented, research studies (Anderson & Helms, 2001; Mendro, 1998; Powell & Anderson, 2002; Strong & Tucker, 2000) cited by Toh et al. (2003) in the journal article titled *Teaching, teacher Knowledge and constructivism* emphasized the pivotal role of teachers.

*You can build new schools, equip them with computers, put up new syllabuses—but the best-laid plans and programmes will fail without your team of competent, dedicated teachers who understand and are committed to the goals set. When all is said and done, it is the teachers who will breathe life into the educational process.*

*(Toh & Tsoi, 2008, p.625)*

**PEDAGOGICAL CONTENT KNOWLEDGE IN THE DIGITAL AGE**

Research evidence tells us that both subject matter knowledge (SMK) and pedagogical content knowledge (PCK) (Shulman, 1986) are crucial to good teaching and student understanding (Reynolds, 1992; Lourdusamy, Toh & Wong, 2001).

PCK is a domain of teacher knowledge that distinguishes the expert teacher in a subject area such as a master science teacher from the subject expert such as a scientist. PCK is that special professional understanding that teachers have whereby they can integrate, transform and represent content knowledge in ways that are comprehensible to the learners. Good teaching is the skillful application of pedagogy for a specific subject matter in particular contexts. This special amalgam of content and pedagogy is a unique class of knowledge that is central to teachers' work that would not be typically held by non-teaching subject matter experts or by teachers who know little of that subject. This blending of content and pedagogy into an understanding of how particular topics, problems or issues are organized, represented, and adapted to diverse interests and abilities of learners, and presented for instruction is known as PCK, as first coined by Shulman.

Conceptually, *PCK* as represented in Figure 1 is an amalgamation of *subject matter knowledge (SMK), general pedagogical knowledge (GPK)* and *contextual knowledge (CtK)*. *SMK* is also commonly known as content knowledge. *GPK* includes classroom management and organization, instructional models and strategies which are ICT-based or non-ICT-based, classroom communication and discourse while *CtK* includes knowledge of learners in a classroom or school and general educational goals and purposes. The key elements in Shulman's conception of PCK are knowledge of representations of subject matter on the one hand and understanding of specific learning difficulties and pupil conceptions on the other hand. These elements are intertwined and should be used in a flexible manner: the more representations teachers have at their disposal and the better they recognize learning difficulties, the more effectively they can deploy their PCK.

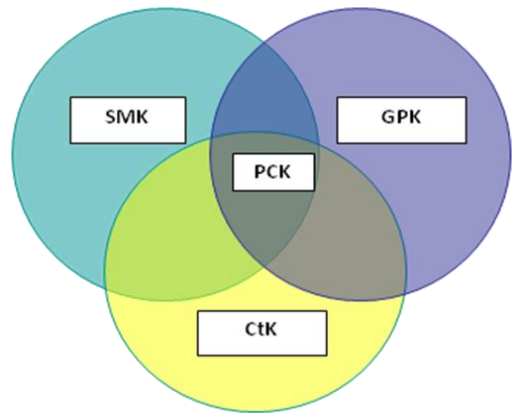

Figure 1. A simplified integrative model of PCK.

**NEED FOR ICT-BASED PEDAGOGICAL MODELS**

Since teaching and learning are two sides of the same coin, teaching in the 21st century must meet the needs of the 21st century learners. Being constantly immersed in an environment where technology is used extensively for information search and social networking, today's students in the digital age expect the use of technology to be part of the lesson design. Today's teachers are increasingly cognizant of the need to integrate digital technologies into their teaching pedagogies as part of their PCK for effective instruction.

While many digitally able teachers can see the potential of digital technology for the purpose of teaching, there is a gap between the digital applications and conceptual understanding of learning and knowledge creation in the digital age (Starkey, 2010). This has led to teachers using digital applications for an array of purposes, with uneven levels of soundness in pedagogy. Due to the explosion of internet application and information, many teachers are overwhelmed with information on ICT tools such as knowledge building software (Wiki, Blogs, Knowledge Forum), concept mapping tools (Cmap tools, Webspiration), communication tools (FaceBook, Twitter) and production tools (YouTube, Glogster), with the wrong assumption that usage of these tools means effective use of ICT in instruction. There is a need for ICT-based pedagogical models that are grounded in both the learning theories of **constructivism** and **connectivism** to guide the teachers' professional practice to engage today's students who are digital natives.

**LEARNING THEORIES OF CONSTRUCTIVISM AND CONNECTIVISM**

As a theory of knowledge, constructivism is founded on the premise that by reflecting on our experiences, we construct our own understanding of the world in which we live. Unlike cognitive constructivism drawn heavily from the work of Piaget's theory of psychological development, social constructivism has its roots in Vygotsky's theory, which applies a sociocultural perspective to psychological development. In social constructivism, social interaction is important as human experiences always include interaction with others. Students want to have their experiential reality confirmed by others and also want to know what others think. Social constructivism argues that learning cannot be separated from the context of learning but rather is a form of cultural apprenticeship and that cognition is situated in specific contexts.

While constructivism has been successful in giving teachers significance to their everyday classroom teaching experience, another learning theory known as connectivism which considers where and how knowledge is created in a digitally enhanced society has emerged. Connectivism which emerged from the notions of complexity thinking, chaos, network and self-organisation theories considers how learning occurs through connecting specialized information sets which can occur between or within organizations, individuals and digital technology. In contrast with constructivism where the focus is on individual learners constructing meaning, connectivism carries the notion of continual expansion of knowledge as new and novel connections open new interpretations and understandings to create new knowledge (Starkey, 2010). Under this theory of connectivism, teachers play the role of the learning experts to facilitate students in the digital age who are seen as co-creators of knowledge via connections in an open and flexible curriculum.

One of the local ICT-based pedagogical models that is grounded in both the learning theories of constructivism and connectivism is the TSOI Hybrid Learning Model (Tsoi, 2007, 2008a, 2008b, 2009a, 2009b, 2010a, 2010b). This hybrid Learning Model is an innovative adaptation from the science learning cycle model (Lawson, 1995) and Kolb's (1984) experiential learning cycle model.

**TSOI HYBRID LEARNING MODEL (HLM): A CASE STUDY**

It is a research evidence-based model which represents learning as a cognitive process in a cycle of four phases: Translating, Sculpting, Operationalizing and Integrating (Figure 2). The unique core of the hybrid learning model is meaningful, functional and relevant for the blended learning experiences. The model represents learning as a cognitive cyclical process of four phases: Translating (experiences translated to a beginning idea of the concept), Sculpting (concept constructed for its critical attributes), Operationalizing (concept internalized for meaningful functionality) and Integrating (concept applied for meaningful transfer of knowledge). Constructivist and inquiry-based, the model advanced from the science learning cycle model (Lawson, 1995) and Kolb's (1984) experiential learning cycle model addresses concept learning and learning styles.

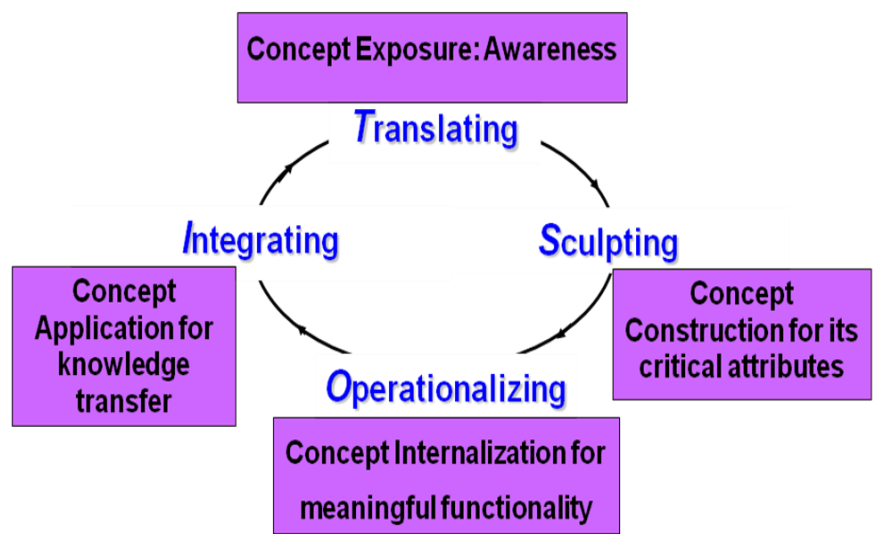

Figure 2. The TSOI Hybrid Learning Model.

Based on the four phases of the TSOI Hybrid Learning Model, a blended learning exemplar on the abstract topic of *Electromagnetic induction* designed by a Physics Task Force comprising a master teacher (Academy of Singapore Teachers), an educational technology officer, an Assistant Professor from the National Institute of Education (NIE) and a scientist from a local data-logger company is illustrated below.

Phase 1 on "Translating" for Concept Exposure

A *face-to-face* demonstration of a real model of producing electricity from magnetism from a hand-held ac generator (Figure 3) provides concept exposure to create awareness in an experiential manner. The experiences are translated to the beginning ideas of the concept.

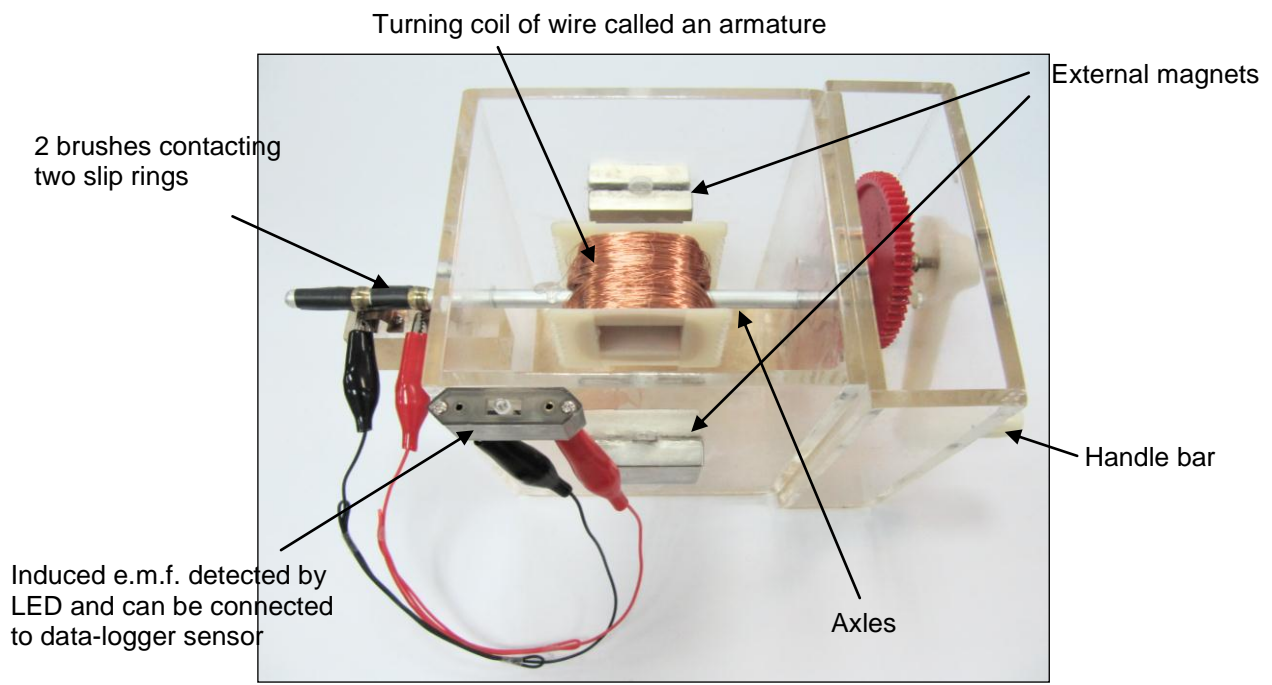

Figure 3. Real-life AC customized demonstration set by a scientist (Dr Tan Kah Chye, 2010)

Phase 2 on "Sculpting" for Concept Construction

A *face-to-face* data-logger based investigation provides concept construction for its critical attributes. Data-loggers (Newton, 2000) are useful tools for the face-to-face components of blended learning as they provide students with both hands-on and minds-on learning experiences with present-day technology. With automatic data collection and graphing, data-logging allows students to spend more time on interpreting data to identify trends and patterns (Sokoloff, Laws, & Thornton, 2007; Thornton & Sokoloff, 1990). Learners will first collect a few sets of data using data-logger (Figure 4), each of which facilitates a preliminary understanding of one aspect of Faraday or Lenz's Laws of Electromagnetic Induction. After a preliminary understanding of Faraday and Lenz's Laws is attained, some crucial characteristics of fairly complex electromagnetic induction graphs obtained under different scenarios lead to conceptual understanding of Faraday and Lenz's Laws.

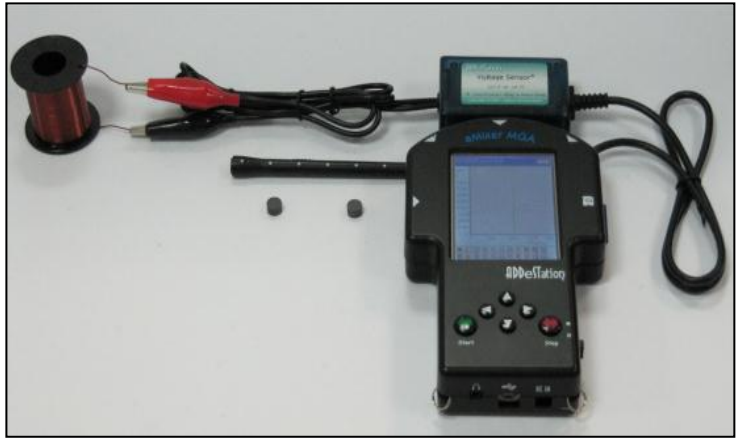

Figure 4: Data-logger apparatus for collection of data

Phase 3 on "Operationalizing" for Concept Internalization

*Online Learning* by students using Ejs Open Source Alternating Current AC Generator Model Java Applet provides concept internalization of Faraday and Lenz's Laws of Electromagnetic Induction leading to meaningful functionality (Figure 5).Through the use of a well-constructed guided inquiry worksheet, learners can engage, explore, explain, elaborate, evaluate and extend learning in the online simulation space at home or in school. Notice this simulation is a close match of the demonstration set (Figure 3). The handle bar, the magnets, 2 slip rings, the armature coil, axle, the output electromotive force (e.m.f.) measured by data-loggers etc. The learner can experience the concepts in the simulations with key advantages here. For example, the rotating of the angle of rotation handle bar $\theta(t)$ cannot be finely controlled in real life; whereas in the simulation, it is possible to input the $\theta(t) = 6.2831*t$ for 1 revolution per second, or any mathematical function such as $\theta(t) = \sin(6.2831*t)$ for rotating clockwise, then anticlockwise or $\theta(t) = t*t$ for increasing angular displacement. This flexibility is limited only by the learners' imagination to explore other scenarios to achieve forms of rotation in real life for concept internalization. Another difficulty to achieve real life example is the visualization of invisible (B) magnetic fields and how it changes when the external magnets exchange position. Visualization of sub-micro electrons moving in the coil wire as the handle bar rotates the coil to generate electromotive force (e.m.f.) can also be demonstrated. Accurate computer models which are well designed and customized can help learners learn concepts more deeply. When these computer models are blended with real life setups, this form of blended learning provides the best of the real physical world and the best of the computer world. Simulations are also useful tools for learning (Belloni, Christian, & Mason, 2009; Hwang & Esquembre, 2003; Nancheva & Stoyanov, 2005; Wieman, Adams, Loeblein, & Perkins, 2010) as students can continue to learn physics at home. This will help overcome the challenges faced by students who cannot continue to interact and learn with physics laboratory apparatus outside school.

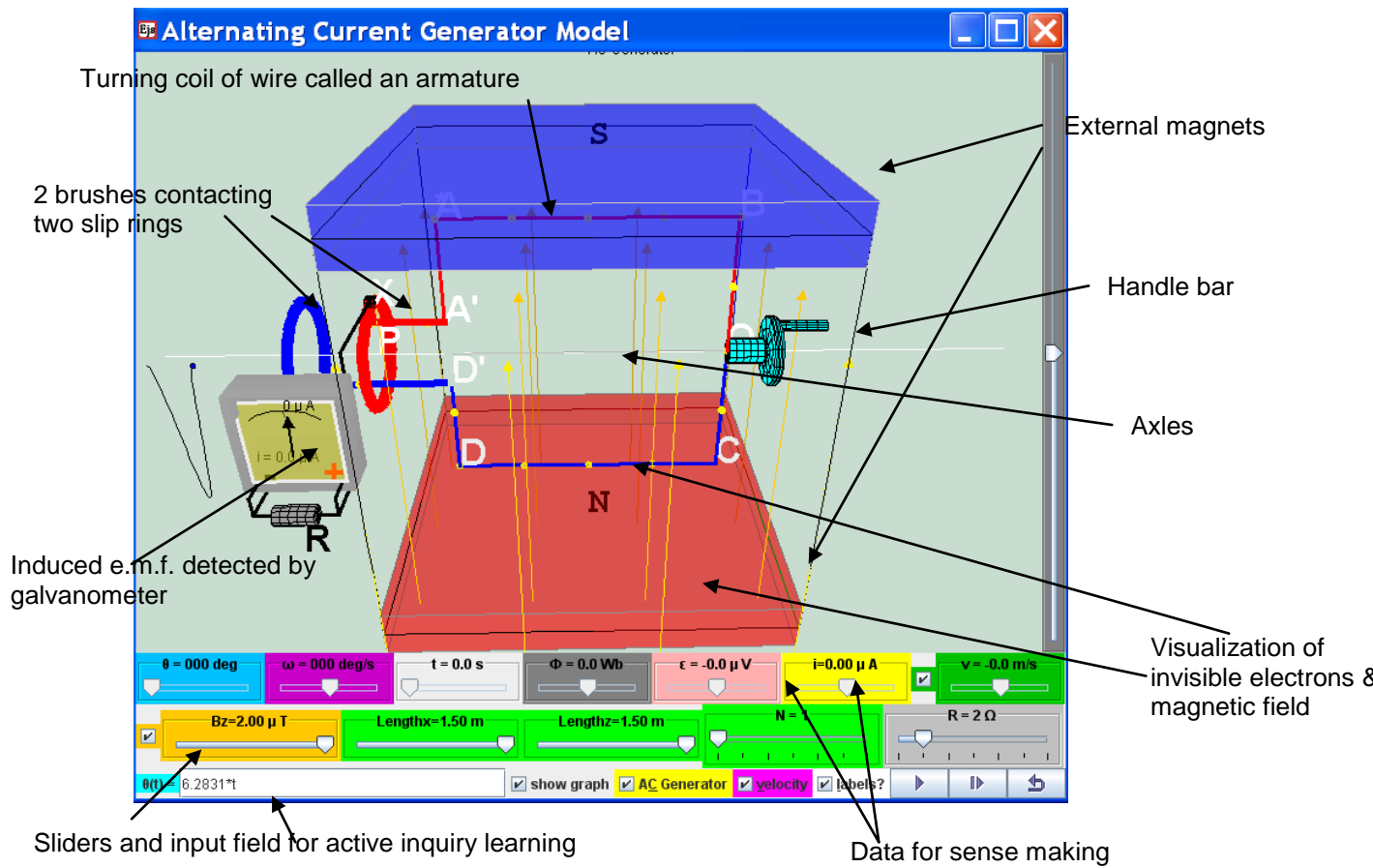

Fig. 5: Screenshot of the AC Generator Simulation Model (Hwang & Wee, 2009)
Source: http://www.phy.ntnu.edu.tw/ntnujava/index.php?topic=1275.msg4924#msg4924

Phase 4 on "Integrating" for Concept Application

A data-logger project featuring *face-to-face* concept application of Faraday and Lenz's Laws to deduce the value of free fall acceleration serves to promote meaningful knowledge transfer. Integrating what is learned and applying to different context is an evidence of good science learning. Learners are to figure out some crucial information with regard to free fall acceleration hidden in an Electromagnetic Induction graph associated with a falling magnet (Figure 6). This helps to exploit the insights into Faraday and Lenz's Laws obtained through Phases 2 and 3. Next, they are to devise a method to deduce the value of free fall acceleration based on the information that they figure out.

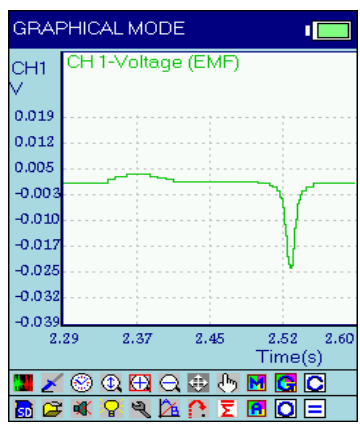

Figure 6. Screenshot of the electromagnetic induction graph associated with falling magnet.

# BLENDED LEARNING FOR 21ST CENTURY TEACHING

The above blended learning exemplar showing the combination of the best elements of face-to-face and online learning is likely to emerge as the predominant mode of teaching and learning to engage today's digital native students who expect their learning environment to include technology. Blended learning is about a mixture of instructional modalities, delivery media, instructional methods, and web-based technologies (Graham, 2006).

Blended learning provides greater access to personalized learning, to resources and experts. Having an online learning component besides face-to-face will allow learners to access learning anywhere and anytime. In addition, the online learning will facilitate a self-paced learning whereby the faster learner can proceed further while the slower learner can proceed at their own pace of learning. The online learning also allows the option of learners staying at home without traveling down to the school, especially in situations such as the severe acute respiratory syndrome (SARS). In addition, blended learning allows greater accommodation for learners and teachers of diverse backgrounds, interests and strengths.

Research has illustrated that it is critical that the methods of delivery match the subject matter knowledge and audience. However, finding one match for everyone is not possible. Instead, a blend of approaches and methods is critical to achieve maximum learning across a variety of learners. A blend of methods and approaches is more likely to produce richer active learning experience and achieve the desired learning outcomes.

Table 1 below shows the blended learning continuum from the traditional face-to-face learning to fully online learning (Watson, 2008).

Table 1. Blended learning continuum (Watson, 2008)

| Fully face-to-Face learning | | | | | | Fully online learning |
|---|---|---|---|---|---|---|
| Traditional face-to-face setting with few or no online resources | Classroom Instruction integrating online resources, but limited or no requirements for students to be online | Classroom Instruction with significant, required online components that extend learning beyond the classroom and beyond the school day | Mostly or fully online curriculum in computer lab or classroom where students meet everyday | Mostly or fully online curriculum with select days required in computer lab or classroom | Fully online curriculum with options for face-to-face instruction, but not required | Fully online curriculum with all learning done online and at a distance and no face-to-face |

In Singapore schools, anecdotal evidence shows that many lessons are currently conducted in the *"Traditional face-to-face settings with few or no online resources"* (column one from the left of Table 1) and "*Classroom Instruction integrating online resources, but limited or no requirements for students to be online*" (column two from the left of Table 1). With the increasing digital literacy of teachers, it is likely that lessons will shift towards "*Classroom Instruction with significant, required online components that extend learning beyond the classroom and beyond the school day*" (third column from the left of Table 1) of the continuum.

**CONCLUSION**

In the book titled *Visible learning:* A synthesis of over 800 meta-analyses relating to achievement (Hattie, 2009), involving 15 years' research on the influences on achievement in school-aged students, Dr John Hattie, Professor of Education and Director of the Visible Learning Labs, University of Auckland, New Zealand, reported the effect sizes of $d = 0.77$ for quality of teaching and $d = 0.74$ for teacher-student relationships. The large effect sizes certainly affirm the importance of quality teaching and teacher-student relationships to bring about meaningful learning.

While blended learning provides pedagogical richness towards active learning for today's students, the pivotal role of the teachers as the interface between the curriculum and the students remains unchanged. The best laid plans and programs can go awry if we do not have teachers who are not only committed to the educational goals of the nation, but also possess a strong pedagogical reasoning and action for the digital age to meet the diverse needs and interests of their students.

**Authors**

**Dr Charles Chew** is currently a Principal Master Teacher (Physics) with the Academy of Singapore Teachers. He has a wide range of teaching experiences and mentors many teachers in Singapore. He is an EXCO member of the Educational Research Association of Singapore (ERAS) and is active in research and publications to promote the importance of theory–practice nexus for reflective instructional practice.

**Loo Kang Wee** is currently a Senior Specialist, Media Design & Technologies for Learning Branch, Education Technology Division at the Ministry of Education. He is a current Physics Subject Chapter Core Team member and was a junior college lecturer. His research is in Open Source Physics tools like Easy Java Simulation for designing computer models and use of Tracker for blended learning.